\documentclass[a4paper,UKenglish,cleveref, autoref, thm-restate]{article}

\usepackage[margin=1.25in]{geometry}

\usepackage{hyperref}

\usepackage[noend]{algpseudocode}
\usepackage{algorithm}
\usepackage{algorithmicx}
\usepackage{dsfont}
\usepackage{todonotes}
\usepackage{amsmath,amsfonts,mathtools}
\usepackage{amsthm}
\usepackage{graphicx}
\usepackage{caption}
\usepackage{subcaption}

\newtheorem{theorem}{Theorem}
\newtheorem{lemma}{Lemma}
\newtheorem{definition}{Definition}
\newtheorem{corollary}{Corollary}

\def\poly{\operatorname{poly}}

\newcommand{\dist}[2]{\ensuremath{\normalfont{\textbf{d}}_{#1}({#2})}}

\title{Near-Optimal Distance Oracles for Vertex-Labeled Planar Graphs} 

\author{Jacob Evald, Viktor Fredslund-Hansen\footnote{Department of Computer Science, University of Copenhagen, \href{mailto:viha@di.ku.dk}{\texttt{viha@di.ku.dk}}. This research is supported by the Starting Grant 7027-00050B from the Independent Research Fund Denmark under the Sapere Aude research career programme.}, Christian Wulff-Nilsen\footnote{Department of Computer Science, University of Copenhagen, \href{mailto:koolooz@di.ku.dk}{\texttt{koolooz@di.ku.dk}}, \href{http://www.diku.dk/koolooz/}{\texttt{http://www.diku.dk/koolooz/}}. This research is supported by the Starting Grant 7027-00050B from the Independent Research Fund Denmark under the Sapere Aude research career programme.}}

\begin{document}

\maketitle

\begin{abstract}
Given an undirected $n$-vertex planar graph $G=(V,E,\omega)$ with non-negative edge weight function $\omega:E\rightarrow \mathbb R$ and given an assigned label to each vertex, a vertex-labeled distance oracle is a data structure which for any query consisting of a vertex $u$ and a label $\lambda$ reports the shortest path distance from $u$ to the nearest vertex with label $\lambda$. We show that if there is a distance oracle for undirected $n$-vertex planar graphs with non-negative edge weights using $s(n)$ space and with query time $q(n)$, then there is a vertex-labeled distance oracle with $\tilde{O}(s(n))$\footnote{We use $\tilde O$-notation to suppress $\poly(\log n)$-factors.} space and $\tilde{O}(q(n))$ query time. Using the state-of-the-art distance oracle of Long and Pettie~\cite{Long2021}, our construction produces a vertex-labeled distance oracle using $n^{1+o(1)}$ space and query time $\tilde O(1)$ at one extreme, $\tilde O(n)$ space and $n^{o(1)}$ query time at the other extreme, as well as such oracles for the full tradeoff between space and query time obtained in their paper. This is the first non-trivial exact vertex-labeled distance oracle for planar graphs and, to our knowledge, for any interesting graph class other than trees.
\end{abstract}

\newpage

\section{Introduction}\label{sec:Intro}
Efficiently answering shortest path distance queries between pairs of vertices in a graph is a fundamental algorithmic problem with a wide range of applications. An algorithm like Dijkstra's can answer such a query in near-linear time in the size of the graph. If we allow for precomputations, we can break this bound, for instance by simply storing the answers to all possible queries in a look-up table. However, a fast query time should preferably not come at the cost of a large space requirement. A \emph{distance oracle} is a compact data structure that can answer a shortest path distance query in constant or close to constant time.

A lot of research has focused on approximate distance oracles which allow for some approximation in the distances output. This is reasonable since there are graphs for which the trivial look-up table approach is the best possible for exact distances. However, for restricted classes of graphs, it may be possible to obtain exact oracles with a much better tradeoff between space and query time. Indeed, for any planar $n$-vertex digraph, there is an exact distance oracle with space close to linear in $n$ and query time close to constant~\cite{Gawrychowski2018,Charalampopoulos2019a,Long2021}.

A related problem is that of obtaining a vertex-labeled distance oracle. Here, we are given a graph with each vertex assigned a label. A query consists of a pair $(u,\lambda)$ of a vertex $u$ and a label $\lambda$ and the output should be the distance from $u$ to the nearest vertex with label $\lambda$. Each vertex is given only one label but the same label may be assigned to multiple vertices. To give some practical motivation, if the graph represents a road network, a label $\lambda$ could represent supermarkets and the output of query $(u,\lambda)$ gives the distance to the nearest supermarket from the location represented by $u$.

Note that this is a generalization of the distance oracle problem since vertex-to-vertex distance queries can be answered by a vertex-labeled distance oracle if each vertex is given its own unique label. If $L$ is the set of labels, a trivial vertex-labeled distance oracle with constant query time is a look-up table that simply stores the answers to all possible queries, requiring space $O(n|L|)$. This bound can be as high as quadratic in $n$.

Our main result, which we shall state formally later in this section, is that for undirected edge-weighted planar graphs, the vertex-labeled distance oracle problem can be reduced to the more restricted distance oracle problem in the sense that up to $\log n$-factors, any space/query time tradeoff for distance oracles also holds for vertex-labeled distance oracles. Hence, the tradeoff from~\cite{Long2021} translates to vertex-labeled distance oracles, assuming that the planar graph is undirected. To the best of our knowledge, this is the first non-trivial upper bound for vertex-labeled distance oracles in any interesting graph class other than trees~\cite{Gawrychowski2018a,Tsur2018}. A strength of our result is that any future progress on distance oracles in undirected planar graphs immediately translates to vertex-labeled distance oracles.

\subsection{Related work on vertex-labeled distance oracles}
Vertex-labeled distance oracles have received considerably more attention in the approximate setting. With $(1+\epsilon)$ multiplicative approximation, it is known how to get $\tilde O(n)$ space and $\tilde O(1)$ query time both for undirected~\cite{Li2013} and directed planar graphs~\cite{Mozes2018} and it has been shown how oracles with such guarantees can be maintained dynamically under label changes to vertices using $\tilde O(1)$ time per vertex relabel.

For general graphs, vertex-labeled distance oracles with constant approximation have been presented~\cite{Hermelin2011,Chechik2012,Gutenberg2018} with state of the art being an oracle with $O(kn|L|^{1/k})$ space, $4k-5$ multiplicative approximation, and $O(\log k)$ query time, for any $k\in\mathbb N$.

\subsection{Our contributions}
We now state our reduction and its corollary:
\begin{theorem}
If there is an exact distance oracle for $n$-vertex undirected edge-weighted planar graphs with $s(n)$ space, $q(n)$ query time, and $t(n)$ preprocessing time, then there exists an exact vertex-labeled distance oracle for such graphs with $s(n) + O(n\log^2 n)$ space, and with $O(q(n)\log n + \log^3 n)$ query time, and $t(n) + \poly(n)$ preprocessing time.
\end{theorem}
Plugging in the distance oracle of Long and Pettie et al.~\cite{Long2021} gives the following corollary which can be seen as a generalization of their result:
\begin{corollary}
For $n$-vertex undirected edge-weighted planar graphs, there exist exact vertex-labeled distance oracles with the following tradeoffs between space and query time:
\begin{enumerate}
    \item $n^{1+o(1)}$ space and $\tilde O(1)$ query time,
    \item $\tilde O(n)$ space and $n^{o(1)}$ query time.
\end{enumerate}
All oracles have preprocessing time polynomial in $n$.
\end{corollary}
Up to logarithmic factors, the full tradeoff between space and query time in their paper similarly extends to vertex-labeled distance oracles in undirected edge-weighted planar graphs.

The rest of the paper is organized as follows. In Section~\ref{sec:defs-notation}, we introduce basic definitions and notation and present tools from the literature that we will need for our oracle. In Section~\ref{sec:vertex-labeled-dist-oracle} we state the key lemmas but defer their proofs until later sections, and thus immediately present our reduction by describing how to obtain a vertex-labeled distance oracle given a distance oracle as a black box. In Section~\ref{sec:point-location}, we present a point location structure similar to~\cite{Gawrychowski2018} but with some important modifications to improve space in our setting.

\section{Preliminaries}\label{sec:defs-notation}
Let $G=(V,E,\omega)$ be a graph with edge weight function $\omega : E \rightarrow \mathbb{R}\cup\left\{ \infty \right\}$. We denote by $V(G) = V$ and $E(G) = E$ the vertex and edge-set of $G$, respectively, and by $n = |V(G)|$ the number of vertices of $G$. A graph $G'$ is said to be a subgraph of $G$ if $V(G') \subseteq V(G)$ and $E(G') \subseteq E(G)$. We denote by $u \leadsto_G v$ a \textit{shortest path} from $u$ to $v$ in $G$, by $\dist{G}{u,v}$ the weight of $u \leadsto_G v$, and write $u \leadsto v = u \leadsto_G v$ and $\dist{}{u,v}=\dist{G}{u,v}$ when $G$ is clear from context. For a shortest path $p = u \leadsto v = (u = p_1),p_2, \hdots, (p_k = v)$ we define vertex $p_i$ to occur \textit{before} $p_j$ on $p$ if $i < j$ and similarly for edges $p_ip_{i+1}$ and $p_{j}p_{j+1}$. Thus statements such as ``the first/last vertex/edge on $p$ satisfying some property $P$'' will always be made w.r.t. this ordering. We also write $p \circ p'$ to denote the concatenation of paths (or edges) $p$ and $p'$, assuming the last vertex of $p$ equals the first vertex of $p'$. Given $u,v,v' \in V$; we say that $v$ is \textit{closer} than $v'$ to $u$ in $G$ if $\dist{G}{u,v} < \dist{G}{u,v'}\text{ or }\dist{G}{u,v} = \dist{G}{u,v}\text{ and }v < v'$, assuming some lexicographic ordering on vertices. We denote by $V(p)$, respectively $E(p)$, the set of vertices, respectively edges, on a path $p$.

Assume in the following that $G$ is undirected. $G$ is said to be connected, respectively biconnected, if any pair of vertices are connected by at least one, respectively two, vertex-disjoint paths. For a rooted spanning tree $T$ in $G$ and for any edge $e = uv$ not in $T$, we define \textit{the fundamental cycle} of $uv$ w.r.t. $T$ as the cycle obtained as the concatenation of $uv$ and the two paths of $T$ from the root to $u$ and $v$, respectively.

\subsection{Planar graphs and embeddings}
An \textit{embedding} of a planar graph $G$ assigns to each vertex a point in the plane and to each edge a simple arc such that its endpoints coincide with those of the points assigned to its vertices. A \textit{planar embedding} of $G$ is an embedding such that no two vertices are assigned the same point and such that no pair of arcs coincide in points other than those corresponding to vertices they share. A graph is said to be \textit{planar} if it admits a planar embedding. When we talk about a planar graph we assume that it is \textit{planar embedded} and hence some implicit, underlying planar embedding of the graph. When it is clear from the context we shall refer interchangeably to a planar graph and its embedding, its edges and arcs and its vertices and points. Thus the term graph can refer to its embedding, an edge to its corresponding arc and a vertex to its corresponding point in the embedding. 

\paragraph*{Assumptions about the input} Unless stated otherwise, we shall always assume that $G$ refers to a graph which is weighted, undirected and planar with some underlying embedding. Furthermore, we shall make the structural assumption that $G$ is triangulated. Triangulation can be achieved by standard techniques, i.e. adding to each face $f$ an artificial vertex and artificial edges from the artificial vertex to each vertex of $V(f)$ with infinite weight. This transformation preserves planarity, shortest paths and ensures that the input graph consists only of simple faces. We also assume that shortest paths in the input graph are unique; this can be ensured for any input graph by either randomly perturbing edge weights or with e.g.~the deterministic approach in~\cite{UniqueShortestPaths} which gives only an $O(1)$-factor overhead in running time. Finally, it will be useful to state the following lemma when talking about separators in a graph with unique shortest paths: 

\begin{lemma}
Let $u,v,x,y \in V(G)$. Then $u \leadsto v$ and $x \leadsto y$ share at most one edge-maximal subpath. \label{lem:sp-join-leave-once}
\begin{proof}
Assume that $x\leadsto y$ intersects $u\leadsto v$ and let $a$ resp.~$b$ be the first resp.~last intersection along $u\leadsto v$. Since $G$ is undirected, uniqueness of shortest paths implies that $a\leadsto b$ is a subpath of $u\leadsto v$ shared by $x\leadsto y$.
\end{proof}
\end{lemma}

\paragraph*{Edge orderings, path turns and path intersections}
For an edge $e = uv$ of a planar embedded graph $H$, we let $<_{e}^{H}$ be the clockwise ordering of edges of $H$ incident to $v$ starting at $e$ (ignoring edge orientations). Hence $<_{e}^{H}$ is a strict total order of these edges and $e$ is the first edge in this order.

For vertices $u,v \in V(H)$, $x \in V(u \leadsto v) \setminus \left\{ u,v \right\}$ and $y \in V(H) \setminus V(u \leadsto v)$, let $pq$ be the last edge shared by $u \leadsto v$ and $x \leadsto y$. Furthermore let $qz$ resp.~$qz'$ be the edge following $pq$ in the traversal of $u \leadsto v$ and $x \leadsto y$, respectively. We say that $x \leadsto y$ \textit{emanates from the left} of $u \leadsto v$ if $qz' <_{pq}^H qz$, and otherwise it \textit{emanates from the right}. We dually say that $y \leadsto x$ \textit{intersects $u \leadsto v$ from the left (right)} if $x \leadsto y$ \textit{emanates from the left (right)}.

Given a face $f$ of $H$, vertices $u \in V(f)$, and $v,v' \in V$, let $H'_f$ be a copy of $H$ with an artificial vertex $f^*$ embedded in the interior of $f$ along with an additional edge $f^*u$. Define the paths $p_v=f^*u \circ u \leadsto_{H'_f} v$ and $p_{v'}=f^*u \circ u \leadsto_{H'_f} v'$ and assume that neither path is a prefix of the other. By assumption and Lemma \ref{lem:sp-join-leave-once}, $p_v$ and $p_{v'}$ share exactly one edge-maximal subpath $f^* \leadsto x$. We say that $u\leadsto_{H} v$ makes a \textit{left turn} w.r.t $u\leadsto_{H} v'$ from $f$ if $x \leadsto v$ emanates from the left of $p_v$, and otherwise it makes a \textit{right turn}; we will omit mention of $f$ when the context is clear. Note that the notion of a turn is symmetric in the sense that $u\leadsto_{H} v$ makes a left turn w.r.t $u\leadsto_{H} v'$ iff $u\leadsto_{H} v'$ makes a right turn w.r.t $u\leadsto_{H} v$.

\subsection{Voronoi Diagrams}
The definitions in this subsection will largely be made in a manner identical to those of~\cite{Gawrychowski2018}, but are included as they are essential to a point location structure which will be presented in Section \ref{sec:point-location}. Given a planar graph $G=(V,E,\omega)$, $S \subseteq V$, the \textit{Voronoi diagram of $S$ in $G$}, denoted by $\textit{VD}(S)$ in $G$ is a partition of $V$ into disjoint sets, $\textit{Vor}(u)$, referred to as \textit{Voronoi cells}, with one such set for each $u \in S$. The set $\textit{Vor}(u)$ is defined to be $\left\{ v \in V \; | \; d(u,v) < d(u',v) \textit{ for all } u' \in S \setminus \left\{ u \right\} \right\}$, that is the set of vertices that are closer to $u$ than any other site in terms of $d(\cdot,\cdot)$. We shall simply write $\textit{VD}$ when the context is clear.

It will also be useful to work with a dual representation of Voronoi diagrams. Let $\textit{VD}^*_0$ be the subgraph of $G^*$ s.t. $E(\textit{VD}^*_0)$ is the subset of edges of $G^*$ where $uv^* \in \textit{VD}^*_0$ iff $u$ and $v$ belong to different Voronoi cells in $\textit{VD}$. Let $\textit{VD}^*_1$ be the graph obtained by repeatedly contracting edges of $\textit{VD}^*_0$ incident to degree 2 vertices until no such vertex remains\footnote{Formally, given a degree 2 vertex $v$ with incident edges $vw, vw'$, we replace these edges by $ww'$, concatenate their arcs and embed $ww'$ using this arc in the embedding.}. We refer to the vertices of $\textit{VD}^*_1$ as \textit{Voronoi vertices}, and each face of the resulting graph $\textit{VD}^*_1$ can be thought of as corresponding to some Voronoi cell in the sense that its edges enclose exactly the vertices of some Voronoi cell in the embedding of the primal. We shall restrict ourself to the case in which all vertices of $S$ lie on a single face $h$. In particular, $h^*$ is a Voronoi vertex, since each site is a vertex on the boundary of $h$ in the primal. Finally, let $\textit{VD}^*$ be the graph obtained by replacing $h^*$ with multiple copies, one for each edge. We note that since there are $|S|$ Voronoi sites (and thus faces in $\textit{VD}^*$), the number of Voronoi vertices in $\textit{VD}^*$ is $O(|S|)$ due to Euler's formula. Furthermore,~\cite{Gawrychowski2018} show that when assuming unique shortest paths and a triangulated input graph, $\textit{VD}^*$ is a ternary tree. It follows that the primal face corresponding to a Voronoi vertex $f^*$ consists of exactly three vertices, each belonging to different Voronoi cells. We refer to the number of sites in a Voronoi diagram as its \textit{complexity}.

Finally, they also note that a \textit{centroid decomposition}, $T^*$, can be computed from $\textit{VD}^*$ s.t. each node of $T^*$ corresponds to a Voronoi vertex $f^*$ and the children of $f^*$ in $T^*$ correspond to the subtrees resulting from splitting the tree at $f^*$, and s.t. the number of vertices of each child is at most a constant fraction of that of the parent. We remark that $\textit{VD}^*(S)$ can be computed by connecting all sites to a super-source and running a single-source shortest paths algorithm, and its centroid decomposition in time proportional to $|V(\textit{VD}^*(S))|$.

\subsection{Separators and decompositions}
In the following, we will outline the graph decomposition framework used by our construction. As part of the preprocessing step, we will recursively partition the input graph using balanced fundamental cycle separators until the resulting graphs are of constant size. We shall associate with the recursive decomposition of $G$ a binary decomposition tree, $\mathcal{T}$, which is a rooted tree whose nodes correspond to the regions of the recursive decomposition of $G$. We will refer to nodes and their corresponding regions interchangeably. The root node of $\mathcal{T}$ corresponds to all of $G$. The following lemma states the invariants of the decomposition that will be used in our construction:

\begin{lemma}\label{lem:decomp-tree}
Let $G=(V,E,\omega)$ be an undirected, planar embedded, edge-weighted, triangulated graph and let $T$ be a spanning tree\footnote{For our purposes, the spanning tree will be a shortest path tree.} of $G$. Then there is an $\tilde{O}(n)$ time algorithm that returns a \textit{binary decomposition tree} $\mathcal{T}$ of $G$ s.t.
\begin{enumerate}
\item for any non-leaf node $G' \in \mathcal{T}$, its children $G'_l$, respectively $G'_r$ corresponds to the non-strict interior, respectively non-strict exterior of some fundamental cycle in $G'$ w.r.t. $T$,
\item for any child node, it contains at most a constant fraction of the faces of its parent,
\item for any leaf node it contains a constant number of faces of $G$,
\item for all nodes at depth $i$, $\mathcal{T}_i$, $\sum_{G' \in \mathcal{T}_i} |V(G')| = O(n)$
\end{enumerate}
\end{lemma}

Properties 1-3 follow from recursively applying a classic linear time algorithm for finding fundamental cycles. Property 4 follows from employing standard techniques that involve contracting degree-two vertices of the separators found at each level of recursion and weighting the resulting edges accordingly. This transformation results in a decomposition where the sum of faces of all regions at any level is preserved. We stress that our construction does not rely on the usual sparse simple cycle separators (of size $O(\sqrt{n})$) but rather fundamental cycle separators of size $O(n)$.

\section{The vertex labeled distance oracle}\label{sec:vertex-labeled-dist-oracle}
In this section we describe our reduction which shows our main result. The reduction can be described assuming Lemma \ref{lem:decomp-tree} and the existence of the point location structure which we will state in the following lemma, the proof of which is deferred to Section~\ref{sec:point-location}:

\begin{lemma}\label{lem:point-location}
Let $G=(V,E,\omega)$ be an undirected, planar embedded, edge-weighted graph with labeling $l : V \rightarrow L$ and let $p$ be a shortest path in $G$. There is a data structure $O_{G,p}$ with $O(|V| \log |V|)$ space which given $u \in V$ and $\lambda \in L$ returns a subset $C \subset V$ of constant size, s.t. if  $v$ is the vertex with label $\lambda$ closest to $u$ and $v \leadsto u$ intersects $p$, then $v \in C$. Each such query takes time at most $O(\log^2 |V|)$.
\end{lemma}

\subsection{Preprocessing} 
Given the input graph $G=(V,E,\omega)$, the preprocessing phase initially computes the decomposition tree, $\mathcal{T}$, of Lemma \ref{lem:decomp-tree}. Associated with each non-leaf node $G' \in \mathcal{T}$ is a fundamental cycle separator of $ab \in E(G')$ w.r.t. the shortest path tree $T$ rooted at some $c \in V(G')$. For such a $G'$ we shall refer to $S_1(G')= c \leadsto_{G'} a$ and $S_2(G') = c \leadsto_{G'} b$. Thus the fundamental cycle separator is given by $S_1(G') \circ ab \circ S_2(G')$. The preprocessing phase proceeds as follows: For all non-leaf nodes $G' \in \mathcal{T}$, compute and store data structures $O_{G',S_1(G')}$ and $O_{G',S_2(G')}$ of Lemma \ref{lem:point-location}. Finally, a distance oracle $D$ with $O(s(|V|))$ space capable of reporting vertex-to-vertex shortest path distances in time $O(t(|V|))$ is computed for $G$ and stored alongside the decomposition tree and the point location structures.

\paragraph*{Space complexity} The decomposition tree $\mathcal{T}$ can be represented with $O(|V| \log |V|)$ space and $D$ with $O(s(n))$ space. For each node $G' \in \mathcal{T}$, we store data structures $S_1(G')$ and $S_2(G')$, so by Lemma \ref{lem:decomp-tree} and \ref{lem:point-location}, we get $$\sum_{i=0}^{\infty} \sum_{G' \in \mathcal{T}_i} |V(G')| \log |V(G')| = \sum_{i=0}^{c \log n} O(|V| \log |V|) = O(|V| \log^2 |V|)$$ for a total space complexity of $O(s(|V|) + |V| \log^2 |V|)$.

\subsection{Query} 
Let $G' \in \mathcal{T}$ and consider the query $\dist{G'}{u,\lambda}$. If $G'$ is a leaf node, the query is resolved in time $O(t(n))$ by querying $D$ once for each vertex of $G'$. If $G'$ is a non-leaf node, the query is handled as follows: First, data structures $O_{G',S_1(G')}$ and $O_{G',S_2(G')}$ are queried with $u$ and $\lambda$, resulting in two ``candidate sets'', $C_1$ and $C_2$, one for each query. By Lemma \ref{lem:point-location}, $C_1 \cup C_2$ contains the nearest vertex with label $\lambda$ for which $u \leadsto_{G'} v'$ intersects either $S_1$ or $S_2$ if such a vertex exists. Compute $d_{G'} = \min \left\{ \dist{G}{u,c} \; | \; c \in C_1 \cup C_2 \right\} \cup \left\{ \infty \right\}$ by querying $D$ once for each vertex of $C_1 \cup C_2$. The query then recursively resolves $d_{G''} = \dist{G''}{u,\lambda}$ where $G''$ is a child of $G'$ in $\mathcal{T}$ containing $u$. Finally, the query returns $\min \left\{ d_{G'}, d_{G''} \right\}$. 

\begin{algorithm}[!hbt]
\caption{Query procedure for the distance oracle.}
\begin{algorithmic}[1]
\Procedure{Query}{$u,\lambda,G'$}
  \If{$G'$ is a leaf node in $\mathcal{T}$}
    \State{\Return{$\min\left\{ \dist{G'}{u,v} \; | \; v \in V(G') \text{ and } l(v) = \lambda \right\}$}}
  \Else
    \State{$C_1 \gets O_{G',S_1(G')}(u,\lambda)$; $C_2 \gets O_{G',S_2(G')}(u,\lambda)$}
    \State{$G'' \gets \text{A child of }G'\text{ in }\mathcal{T}\text{ containing $u$}$}
    \State{$d_{G'} \gets \min \left\{ \dist{G}{u,c} \; | \; c \in C_1 \cup C_2 \right\} \cup \left\{ \infty \right\}$}
    \State{$d_{G''} \gets \textsc{Query}(u,\lambda,G'')$}
    \State{\Return{$\min \left\{ d_{G'}, d_{G''} \right\}$}}
 \EndIf
\EndProcedure
\end{algorithmic}
\label{alg:final-query}
\end{algorithm}

\paragraph*{Correctness} Denote by $v$ the vertex of $G$ with label $\lambda$ nearest to $u$ in $G'$, and consider the case in which $u \leadsto_{G'} v$ intersects $S_1(G')$ or $S_2(G')$. In this case, $v \in C$ by Lemma \ref{lem:point-location} and $C \neq \emptyset$, so $$d_{G'} = \min \left\{ \dist{G}{u,c} \; | \; c \in C \right\} = \dist{G}{u,v} = \dist{G'}{u,v}  = \dist{G'}{u,\lambda} \leq d_{G''}$$
with the inequality following from definition of $v$. Note that in case $u$ is a vertex of either $S_1$ or $S_2$, the correct estimate is returned at the current level, but for a simpler description, the recursion proceeds anyways. Otherwise $u \leadsto_G' v$ intersects neither $S_1$ and $S_2$ in which case, the path must be fully contained in the (unique) child node, $G''$, of $G'$ containing $u$. In this case, the query reports $d_{G''} = \dist{G''}{u,\lambda} = \dist{G''}{u,v} \leq d_{G'}$, showing the correctness.

\paragraph*{Time complexity} At each level of the recursion, $O_{G',S_1(G')}$ and $O_{G',S_2(G')}$ are queried in time $O(\log^2 |V|)$. Furthermore, $D$ is queried $|C| = O(1)$ times in total time $O(1) \cdot O(t(|V|)) = O(t(|V|))$. By Lemma \ref{lem:decomp-tree}, the query is recursively resolved on a problem instance which is a constant fraction smaller at each level of recursion, giving rise to the recurrence relation $T(n) = T(n/a) + O(t(n) + \log^2 n)$. When $G'$ is a leaf node, then by Lemma \ref{lem:decomp-tree}, $G'$ consists of a constant number of faces, described by the base case $T(n) = O(t(n))$ when $n \leq b$ for sufficiently small $b$. It is easily verified that a solution to the recurrence is bounded by $O(\log^3 n +  t(n)\log n)$. This shows the main theorem, and the rest of this paper is devoted to proving Lemma \ref{lem:point-location}.

\section{The point location data structure} \label{sec:point-location}
Our point-location data-structure uses techniques similar to those of~\cite{Gawrychowski2018} for point location in additively weighted Voronoi diagrams, but with some crucial differences in order to save space.

Both structures rely on being able to determine left/right turns of shortest paths in shortest path trees rooted at sites in $G$, but to facilitate this, the data structure of~\cite{Gawrychowski2018} explicitly stores an (augmented) shortest path tree rooted at each site as well as a data structure for answering least common ancestor (LCA) queries. The point location structure thus requires $\Theta(|S|n)$ space where $S$ is the number of sites, and since $S$ may be large as $\Theta(\sqrt{n})$ (corresponding to the size of a sparse balanced separator in a planar graph), this translates to $\Theta(n^{3/2})$ space for their problem. This will not work in our case since the number of sites can in fact be as high as $\Theta(n)$, leading to a quadratic space bound.

The second issue with applying the techniques from~\cite{Gawrychowski2018} directly to our setting, is that it requires us to store a Voronoi diagram for each label, for each shortest path. Each vertex of the path separator would then be a site in the stored Voronoi diagram but as each separator may be large, i.e. $\Theta(n)$, we may use as much as $\Theta(n|L|)$ space over \emph{all} labels of $L$ for a single separator. What we need is for the number of sites involved for a label $\lambda$ to be proportional to the number of vertices with label $\lambda$; this would give a near-linear bound on the number of sites when summing over all $\lambda\in L$ across all levels of the recursive decomposition. We address these issues in the following sections.

\subsection{MSSP structure}
To compactly represent shortest path trees, our point location structure uses an augmented version of the multiple-source shortest-path (MSSP) data structure of Klein~\cite{Klein2007}. It cleverly uses the persistence techniques of~\cite{Driscoll1989} in conjunction with top trees~\cite{Holm2003} to obtain an implicit representation of shortest path trees rooted at each site. Top-trees allow for shortest path distance queries and least-common ancestor (LCA) queries in time $O(\log n)$ per query while using $O(n \log n)$ space, and can easily be augmented to support turn queries, as we shall see shortly. To be used as a black box, the MSSP structure relies on being initialized from a face of $G$. In our construction, we wish to use it for querying left/right turns of paths and distances from vertices residing on shortest paths of fundamental cycle separators, and thus some further preprocessing is required. The guarantees of the augmented MSSP structure used for the point location structure are summarized in the following lemma:

\begin{lemma}\label{lem:klein-enhanced}
Let $G=(V,E,\omega)$ be an edge-weighted planar graph, $f$ be a face of $G$ and let $T_u$ denote the shortest path tree rooted at $u$. Then there exists a data structure $\text{MSSP}(G,f)$ with $O(n \log n)$ space which given $u \in V(f)$ and $c,v \in V$ supports queries
\begin{enumerate}
\item $\textsc{Dist}(u,v)$: report $\dist{G}{u,v}$,
\item $\textsc{LCA}(u,c,v)$: report the least common ancestor of $v,c$ in $T_u$,
\item $\textsc{Turn}(u,c,v)$: report whether $u \leadsto_{T_u} c$ makes a left or right turn w.r.t. $u \leadsto_{T_u} v$ or if one is a prefix of the other.
\end{enumerate}
in time $O(\log n)$ per query. The data structure can be preprocessed in $O(n \log n)$ time.
\end{lemma}

Descriptions of $\textsc{Dist}$ and $\textsc{LCA}$ are available in~\cite{Klein2007} and~\cite{Holm2003}, and a description of how to implement $\textsc{Turn}$ is provided in Appendix~\ref{sec:appendix} in terms of the vocabulary and interface specified in~\cite{Holm2003} for completeness. A top-tree representing any shortest path tree rooted at a vertex on $f$ can be accessed in time $O(\log n)$ by  using persistence in the MSSP structure. Lemma \ref{lem:klein-enhanced} then readily follows from applying the bound of Lemma \ref{lem:turn} in Appendix~\ref{sec:appendix}.

\subsection{Label sequences}
To address the second issue, we first need to make the following definition and state some of its properties when applied in the context of planar graphs:

\begin{definition}\label{def:interval}
Let $G=(V,E)$ be a graph, $p=p_1,\hdots,p_k$ a sequence of vertices and $S \subseteq V$. The \textit{label-sequence of $p$ w.r.t. $S$} is a sequence $M_{G,S,p} \in S^k$ satisfying $M_{G,S,p}(i) = \arg\min_{s \in S} \text{dist}_G(s,p_i)$. The \textit{alternation number} on $p$ w.r.t. $S$ in $G$ is defined as $|M_{G,S,p}| = \sum_{i=1}^{k-1} [M_{G,S,p}(i) \neq M_{G,S,p}(i+1)]$.
\end{definition}
When $G$, $S$ and $p$ are clear from the context, we shall simply write $M$, and also note that the sequence is well-defined due the tie-breaking scheme chosen in the preliminaries. The alternation number can be thought of as the number of times consecutive vertices on $p$ change which vertex they are closest to among $S$ when ``moving along'' $p$.

When $G$ is an undirected planar graph and $p$ is a shortest path in $G$, it can be observed\footnote{We thank the anonymous reviewer for this observation which saved a tedious proof.} that $M$ is essentially a Davenport-Schinzel sequence of order 2, and it immediately follows that the alternation number is ``small'' in the sense it is proportional to $S$ while being agnostic towards the length of $p$ altogether.

\begin{definition}[Davenport-Schinzel~\cite{Davenport1965}]\label{def:davenport}
A sequence $u_1, u_2, \hdots, u_k$ over an alphabet $\Sigma$ on $n$ symbols is said to be a $(n,s)$-Davenport-Schinzel sequence if
\begin{enumerate}
\item $u_i \neq u_{i+1}$ for all $1 \leq i < k$ and
\item There do not exist $s+2$ indices $1 \leq  i_1 < i_2 < \hdots < i_{s+2} \leq k$ for which $u_{i_1}=u_{i_3} = \hdots = u_{i_{s+1}} = u$ and $u_{i_2}=u_{i_4} = \hdots = u_{i_{s+2}} = v$ for  $u \neq v \in \Sigma$.
\end{enumerate}
\end{definition}

\begin{lemma}[Davenport-Schinzel~\cite{Davenport1965}]\label{lem:davenport}
Let $U$ be a $(n,2)$-Davenport-Schinzel sequence of length $m$. Then $|U| \leq 2n-1$.
\end{lemma}

For a sequence of $\mathcal S = u_1,u_2,\ldots,u_k$ over an alphabet $\Sigma$, the \emph{contraction} of $\mathcal S$ is the subsequence obtained from $\mathcal S$ by replacing every maximal substring $s,s,\ldots, s$ of $\mathcal S$ consisting of identical symbols $s$ by a single occurence of $s$. As an example with $\Sigma = \{0,1,2\}$, the contraction of $ 0,0,1,2,2,1,1,0,1,0,0,2$ is $0,1,2,1,0,1,0,2$.

\begin{figure}
\centering
\begin{subfigure}{.45\textwidth}
  \centering
  \includegraphics[page=6,width=.9\linewidth]{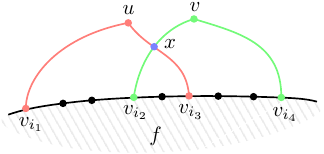}
  \caption{Vertex $v$ not contained in $C$.}
  \label{fig:dp-cases1}
\end{subfigure}%
\begin{subfigure}{.45\textwidth}
  \centering
  \includegraphics[page=7,width=.9\linewidth]{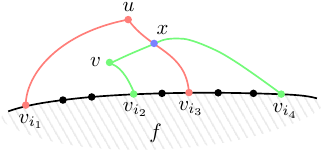}
  \caption{Vertex $v$ contained in $C$.}
  \label{fig:dp-cases2}
\end{subfigure}
\caption{Illustration of the proof of Lemma \ref{lem:alternation}. The concatenation of $u \leadsto v_{i_1}$, $v_{i_1}\leadsto v_{i_3}$, and the reverse of $u\leadsto v_{i_3}$ forms a cycle $C$ and $v \leadsto v_{i_2}$ intersects $u \leadsto v_{i_3}$ in $x$. Note that w.l.o.g. $C$ is not necessarily simple, that $v \leadsto v_{i_2}$ may intersect the cycle more than once and that $v$ may be a vertex of $C$.}
\label{fig:dp-cases}
\end{figure}

\begin{lemma}\label{lem:alternation}
Let $G$ be an undirected, weighted planar graph, let $S \subseteq V$, and let $p$ be a shortest path of $G$ contained in (the boundary of) a face of $G$. Then the contraction of $M$ is a  $(|S|,2)$-Davenport-Schinzel sequence.
\begin{proof}
Define $v_1, \hdots, v_k = p$ and assume for the sake of contradiction that for some $1 \leq i_1 < i_2 < i_3 < i_4 \leq k$ and $u,v \in S$ with $u\neq v$, it holds that $u = M(i_1) = M(i_3)$ and $v = M(i_2) = M(i_4)$.  Then the concatenation of $u \leadsto v_{i_1}$, $v_{i_1}\leadsto v_{i_3}$, and the reverse of $u\leadsto v_{i_3}$ forms a cycle. Thus, either $v\leadsto v_{i_2}$ intersects $u\leadsto v_{i_1}\cup u\leadsto v_{i_3}$ or $v\leadsto v_{i_4}$ intersects $u\leadsto v_{i_1}\cup u\leadsto v_{i_3}$; see Figure \ref{fig:dp-cases1} and \ref{fig:dp-cases2}. By symmetry, we only need to consider the former case. If $v\leadsto v_{i_2}$ intersects $u\leadsto v_{i_3}$ in some vertex $x$ then $v\leadsto x$ has the same weight as $u\leadsto x$. By the ``closer than''-relation, $u = M(i_3) = M(i_2) = v$, contradicting our assumption that $u\neq v$. A similar contradiction is obtained if $v\leadsto v_{i_2}$ intersects $u\leadsto v_{i_1}$.
\end{proof}
\end{lemma}

\begin{corollary}\label{cor:interval-bound}
Let $G$, $S$ and $p$ be as in Lemma \ref{lem:alternation}. Then $|M_{G,S,p}|=O(|S|)$.
\end{corollary}

We remark that $M$ can be readily computed in polynomial time by adding a super-source connected to each vertex of $S$ and running an SSSP algorithm. Furthermore $M$ can be represented with $O(S)$ space, by storing only the indices for which $M(i) \neq M(i+1)$ and $M(i)$ for each such index. 

We will now describe how to achieve $O(n)$ space for storing Voronoi diagrams for all labels $\lambda \in L$ at any level of the recursive decomposition. We do so by modifying the preprocessing steps and query scheme of \cite{Gawrychowski2018} in a manner suitable for application of Lemma \ref{lem:alternation} and Corollary \ref{cor:interval-bound}.

\subsection{Preprocessing}
Let us briefly recall the statement of Lemma \ref{lem:point-location}; that is, we let $G$ be an undirected, edge-weighted, planar embedded graph with associated labeling $l : V \rightarrow L$ and let $p=p_1 \leadsto p_k$ be a shortest path in $G$. Given a query $(u,\lambda) \in V \times L$, we want to identify a small ``candidate'' set of vertices $C \subseteq V$ such that if $v$ is the vertex with label $\lambda$ closest to $u$ and $u \leadsto v$ intersects $p$, then $v \in C$.

Here, we first describe how to compute a data structure which provides the guarantees of Lemma \ref{lem:point-location}, but restricts itself to the case only where $u \leadsto v$ intersects $p$ \textit{from the left}. The description of the data structure for handling paths that intersect $p$ from the right is symmetric (e.g. by swapping the endpoints of $p$). Lemma \ref{lem:point-location} thus readily follows from the existence of such structures. 

\begin{figure}
\centering
\begin{subfigure}{.49\textwidth}
  \centering
  \includegraphics[page=1,width=.9\linewidth]{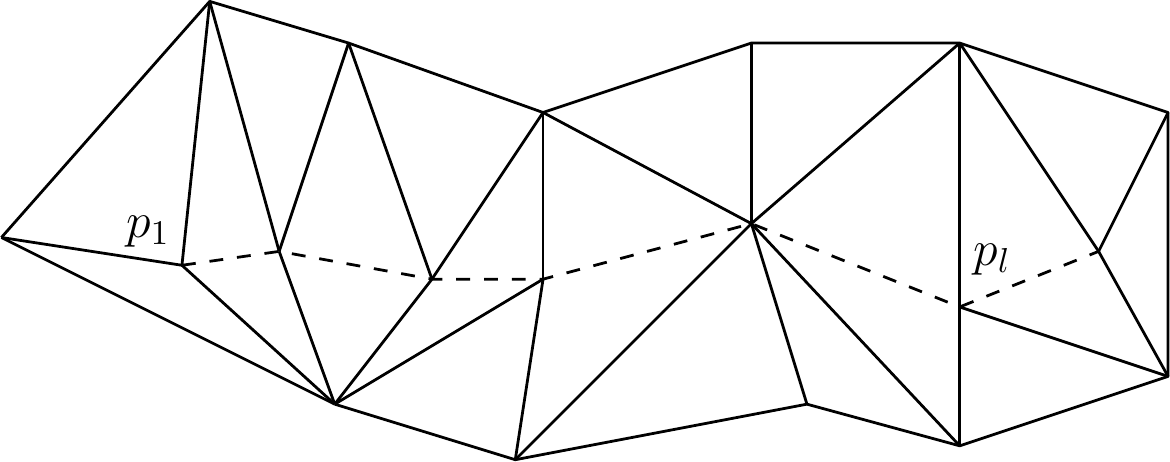}
  \caption{$G$ before incision. Here $p$ as indicated by the dashed line.}
  \label{fig:incision}
\end{subfigure}%
\hfill
\begin{subfigure}{.49\textwidth}
  \centering
  \includegraphics[page=2,width=.9\linewidth]{images/images.pdf}
  \caption{The resulting graph $G_p$ after the incision; $p$ is replaced by two paths $p'$ and $p''$ that enclose the face $f_p$.}
  \label{fig:incision2}
\end{subfigure}
\vfill
\begin{subfigure}{.49\textwidth}
  \centering
  \includegraphics[page=3,width=.9\linewidth]{images/images.pdf}
  \caption{The face $f'_p$ resulting from adding edges to $r_i$. Here $b_1=a_2$ and $b_2=a_3$}
  \label{fig:fp}
\end{subfigure}
\hfill
\begin{subfigure}{.49\textwidth}
  \centering
  \includegraphics[page=4,width=.9\linewidth]{images/images.pdf}
  \caption{The Voronoi diagram is represented by a colored shortest-path tree for each site $r_i$.}
  \label{fig:voronoi}
\end{subfigure}
\caption{Preprocessing steps for the point-location structure.}
\end{figure}

First, a copy, $G_p$, of $G$ is stored and an incision is added along $p$ in $G_p$. This results in a planar embedded graph $G_p$, which has exactly one more face than $G$. Define by $p'=p'_1, \hdots, p'_l$ and $p''=p''_1, \hdots, p''_l$ the resulting paths along the incision, where $p'_1 = p''_1$ and $p'_l = p''_l$. We denote by $f_p$ the face whose boundary vertices are $V(p') \cup V(p'')$. An illustration of this is provided in Figure \ref{fig:incision} and \ref{fig:incision2}.

Next, the MSSP data structure of Lemma \ref{lem:klein-enhanced}, $\textit{MSSP}(G_p,f_p)$, is computed and stored as part of the point-location data structure. This structure will be used for the point location query.

\paragraph*{Centroid decompositions of Voronoi diagrams} The following preprocessing is now done for \textit{each} label $\lambda \in L$: First a copy, $G^\lambda_p$, of $G_p$ is made. Next, $M_{G_p,S_\lambda, p'}$ is computed for $S_\lambda = \left\{ v \in V \; | \; l(v) = \lambda \right\}$. For convenience we assume that $M(0) = \textsc{nil}$. The preprocessing phase now consists of modifying $G^\lambda_p$ before computing the Voronoi diagram and the associated centroid decomposition associated with $\lambda$ as follows: For $i \gets 1, \hdots, l$, whenever $M(i) \neq M(i-1)$, a new vertex is added to $G^\lambda_p$ and embedded in $f_p$ along the curve formed by the deleted arc of the embedding of $p$. Denoting by $r_i$ the most recently added vertex after iteration $i$, edge $p'_ir_i$ with $\omega(p'_ir_i) = d_G(M(i),p'_i)$ is added to $G^\lambda_p$ and embedded for all $i$. Once again, it is fairly easy to see that $G^\lambda_p$ is planar embedded. See Figure \ref{fig:fp} for an illustration of this. Denote by $a_i$ and $b_i$ the endpoints of the first and last edges added incident to $r_i$ (w.r.t. the order in which they were added). Denote by $f'_p$ that has $\left\{ r_i, a_i, b_i\; | \; 1 \leq i \leq l \right\} \cup V(p'')$ as its boundary vertices. Now the Voronoi diagram, its dual and subsequently its corresponding centroid decomposition, $T^*_{p,\lambda}$, is computed in (the now modified) $G^\lambda_p$ using $R = \left\{ r_i \; | \; 1 \leq i \leq l \right\}$ as Voronoi sites, see Figure \ref{fig:voronoi}. The intuition is that each site in $R$ corresponds to a contiguous subsequence $M(k),\hdots,M(l)$ of $M$ for which $M(j)=M(j+1)=v$ where $v$ is the vertex with label $\lambda$ closest to $p'_j$ for $k \leq j < l$. This implies that the number of sites is proportional to the number of vertices with label $\lambda$ instead of the length of the original separator $p$: By Lemma \ref{cor:interval-bound}, $|M| = O(|S_\lambda|)$ and since $|R| = |M|$ it follows that $|R| = O(|S_\lambda|)$ bounds the complexity of $T^*_{p,\lambda}$, which is stored as part of the data structure. As aforementioned, each centroid $c \in T^*_{p,\lambda}$ corresponds to some degree three Voronoi vertex, $f_c^*$, with vertices, $\left\{ x_1,x_2,x_3 \right\}$ in the corresponding primal face $f_c$ s.t. each $x_j$ belongs to a different Voronoi site $r_{i_j}$ for $j \in \left\{ 1,2,3 \right\}$. For each such $j$, the centroid $c$ stores a pointer to its corresponding face $f_c$, the first vertex $p_{k_j}$ of $p'$ on $r_{i_j} \leadsto_{G_p'} x_{j}$ and the weight $\omega(p_{k_j}r_{i_j})$.

\paragraph*{Space complexity}
The space used for storing the MSSP structure is $O(|V| \log |V|)$ by Lemma \ref{lem:klein-enhanced}. For each centroid, we store a constant amount of data, so the space required for storing the centroid decompositions is
$$\sum_{\lambda \in L} O(1) \cdot |T^*_{p',\lambda}| = \sum_{\lambda \in L} O(|S_\lambda|) = O(V)$$
since $S_\lambda \cap S_{\lambda'} = \emptyset$ for $\lambda \neq \lambda' \in L$ as each vertex has exactly one label. The total space used is thus $O(|V| \log |V|)$.

\subsection{Handling a point location query}
We now show how to handle a point location query. Note that in the following, we can assume that the vertices of $p'$ appear in increasing order of their indices when traversing the boundary of $f_p$ in a clockwise direction. Recall that given $u \in V$ and $\lambda \in L$, we wish to find a subset $C \subset V$ of constant size, s.t. if $v$ is the closest vertex with label $\lambda$ where $u \leadsto v$ intersects $p$ from the left, then $v \in C$. The query works by identifying a subset $P \subseteq V(p')$ s.t. for some $p'_k \in P$ it holds that $M_{G_p,S_\lambda,p'}(k) = v$. We first show how to identify the subset by recursively querying the centroid decomposition $T^*_{p,\lambda}$ according to the following lemma, which we note is modified version of the query in \cite{Gawrychowski2018}:

\begin{lemma}\label{lem:point-location-inner}
Given a query $(u,\lambda) \in V\times L$, consider the centroid decomposition tree $T^*_{p,\lambda}$ computed from $G^\lambda_p$ in the preprocessing. Let $c$ be a centroid $c \in T^*_{p,\lambda}$ corresponding to some Voronoi vertex, $f_c^*$, with associated primal triangle containing vertices $\left\{ x_0,x_1,x_2 \right\} = V(f_c)$ where $x_j$ belongs to the Voronoi cell of $r_{i_j}$ for $j \in \left\{ 0,1,2 \right\}$ and $i_0 < i_1 < i_2$. Furthermore let $e^*_{j}$ be the dual edge incident to $f^*_c$, s.t. $e_{j} = x_{j}x_{(j+1) \text{ mod }3}$, let $p_{k_j}$ be the successor of $r_{i_j}$ on $r_{i_j} \leadsto_{G_p^\lambda} x_j$, let $P_j = p_{k_j} \leadsto_{G_p} u$, and let $T^*_j$ be the subtree of $T^*_{p,\lambda}$ attached to $c$ by $e^*_j$ for $j \in \{ 0,1,2 \}$. Finally, let $j^* = \arg\min_{j \in \{ 0,1,2 \}}\{ \dist{G_p}{p_{k_j},u} + \omega(r_{i_j}p_{k_j}) \} = \arg\min_{j \in \{ 0,1,2 \}}\{ \dist{G_p^\lambda}{r_{i_j},u} \}$. Then 
\begin{enumerate}
\item If $p_{k_{j^*}}\leadsto_{G_p} u$ emanates from the left of $P_{j^*}$ or $u\in P_{j^*}$, then the site closest to $u$ in $G_p^\lambda$ belongs to $R^- = \{r_{(i_{(j^*-1)\mod 3}},\ldots,r_{i_{j^*}}\}$ and the second vertex on the shortest path from that site to $u$ in $G_p^\lambda$ belongs to $P^- = \{p_{(i_{(j^*-1)\mod 3}},\ldots,p_{i_{j^*}}\}$; furthermore, $T^*_{j^*}$ is the centroid decomposition tree for $G_p^\lambda$ when restricted to shortest paths from sites in $R^-$ through successors in $P^-$. 
\item otherwise, the site closest to $u$ in $G_p^\lambda$ belongs to $R^+ = \{r_{i_{j^*}},\ldots,r_{i_{(j^*+1)\mod 3}}\}$ and the second vertex on the shortest path from that site to $u$ belongs to $P^+ = \{p_{i_{j^*}},\ldots,p_{i_{(j^*+1)\mod 3}}\}$;  furthermore, $T^*_{(j^*+1)\mod 3}$ is the centroid decomposition tree for $G_p^\lambda$ when restricted to shortest paths from sites in $R^+$ through successors in $P^+$. 
\end{enumerate}
\begin{proof}
By symmetry, we only consider the first case since the second case occurs when $p_{k_{j^*}}\leadsto_{G_p} u$ emanates from the right of $P_{j^*}$ and the first case also includes the case where $u\in P_{j^*}$.

By the choice of $j^*$, $p_{k_{j^*}}\leadsto_{G_p} u$ cannot intersect any of the paths $P_{j'}$ with $j'\in\{(j^*-1)\mod 3,(j^*+1)\mod 3\}$ since these two paths are subpaths of shortest paths from sites $r_{i_{j'}}\neq r_{i_{j^*}}$ in $G_p^\lambda$ and we assume unique shortest paths. Let $x$ be the first vertex of $p_{k_{j^*}}\leadsto_{G_p} u$ such that either $x = u$ or the path emanates from the left of $P_{j^*}$ at $x$. The rest of $p_{k_{j^*}}\leadsto_{G_p} u$ following $x$ will not intersect $P_{j^*}$ again due to uniqueness of shortest paths. Thus $u$ belongs to the region of the plane enclosed by paths $P_{(j^*-1)\mod 3}$, $P_{j^*}$, edge $e_{(j^*-1)\mod 3}$, and path $p_{k_{(j^*-1)\mod 3}},\ldots,p_{k_{j^*}}$. Note that $T^*_{j^*}$ is the subtree of $T^*_{p',\lambda}$ spanning the primal faces contained in this region. Hence, $T^*_{j^*}$ is the centroid decomposition tree for $G_p^\lambda$ when restricted to shortest paths from sites in $R^-$ through successors in $P^-$. 
\end{proof}
\end{lemma}

\begin{figure}
    \centering
    \includegraphics[page=5,width=1.0\textwidth]{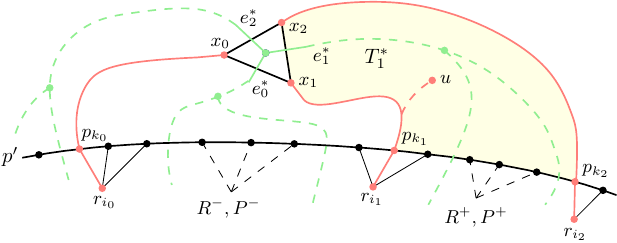}
    \caption{Illustration of Lemma \ref{lem:point-location-inner}. The dashed green lines represent Voronoi edges in the centroid decomposition and the red lines the primal shortest paths to the primal vertices of the centroid. In this case, $j^* = 1$, so $u$ is contained in primal faces spanned by the subtree $T^*_{1}$ contained in the region highlighted in yellow. }
    \label{fig:point-location}
\end{figure}

For an illustration of Lemma \ref{lem:point-location-inner}, see Figure \ref{fig:point-location}. The Lemma implies a fast recursive point location scheme. On query $(u,\lambda)$, obtain centroid $c$ from $T^*_{p',\lambda}$. Since weights of edges from sites have been precomputed, $\textit{MSSP}(G_p,f_p)$ is applied to find $j^*$. $\textit{MSSP}(G_p,f_p)$ is also used to determine if $p_{k_{j^*}}\leadsto_{G_p} u$ emanates from the left of $P_{j^*}$ and hence whether the first or second case of the lemma applies. The point location structure now recurses on a subset of sites and vertices of $p'$ and on a subtree of $T^*_{p',\lambda}$, depending on which case applies.

The recursion stops when reaching a subtree corresponding to a bisector for two sites. The vertices of $V$ with label $\lambda$ corresponding to these two sites are reported as the set $C$, yielding the desired bound. 

\paragraph*{Time complexity}
The $O(\log^2|V|)$ query time bound of Lemma~\ref{lem:point-location} follows since there are $O(\log|V|)$ recursion levels and in each step, a constant number of queries to $\textit{MSSP}(G_p,f_p)$ are executed, each taking $O(\log |V|)$ time.

\section{Acknowledgments}
We are grateful for the thorough and useful feedback provided by the reviewers.

\newpage
\bibliography{paper}
\newpage

\section*{Appendix A}\label{sec:appendix}
A description of how to implement $\textsc{Turn}$ in Lemma \ref{lem:klein-enhanced} is provided here in terms of the terminology and the interface specified in~\cite{Holm2003}. Readers that are not familiar with the terminology and interface pertaining to top-trees are referred to~\cite{Holm2003}.

\begin{lemma}\label{lem:turn}
Let $G=(V,E,\omega)$ be a weighted planar graph, and let $\mathcal{T}$ be the root-cluster of a top-tree corresponding to some tree $T$ in $G$. Then in addition to $\textsc{Dist}$ and $\textsc{LCA}$, we can support $\textsc{Turn}$ queries: For $u,c,v \in V$, report whether $u \leadsto c$ makes a left or right turn w.r.t. $u \leadsto_T v$, or if one is a prefix of the other in time $O(\log n)$ per query.
\begin{proof}
A description of how to perform $\textsc{LCA}$ queries is found in \cite{Holm2003}. Assume that $u,c,v \in T$ and w.l.o.g. that $u$ is strictly more rootward in $T$ than $v$. First use $\mathcal{T}$ to determine the LCA $c'$ of $(v,c)$. If $c'$ is on both $u \leadsto c$ and $u \leadsto v$ one path is a prefix of the other. Otherwise invoke $\textsc{expose}(u,c')$ and traverse $\mathcal{T}$ until a leaf of $\mathcal{T}$ corresponding to the edge, $e_{v} \in E(T)$ is reached, which connects $c'$ to the subtree containing $v$ in $T$. This can be done in time $O(\log n)$. The same is done for $(u,c')$ and $(c,c')$, exposing edges $e_u, e_c \in T$. Now, if $e_c = e_v$ or $e_c = e_u$, $c$ must be on $u \leadsto_T v$. Otherwise it is easily checked, by maintaining a cyclic order of edges in the adjacency list of $c'$, in constant time, whether $e_c$ emanates to the left or right of the subpath $e_ue_v$ and hence $u \leadsto_T v$. The total time spent is $O(\log n)$.
\end{proof}
\end{lemma}

\end{document}